\documentclass[preprint,aps,prc,superscriptaddress,nofootinbib,amsmath]{revtex4}

\usepackage{graphicx}
\usepackage{bm}
\usepackage{dcolumn}
\usepackage{longtable,multirow}
\usepackage{rotating}
\usepackage{subfigure}
\begin{document}

\title{Thermodynamic Properties of $^{56,57}$Fe}

\author{E.~Algin} 
\affiliation{Department of Physics, Eskisehir Osmangazi University, Meselik, 26480 Turkey}
\author{U.~Agvaanluvsan}
\affiliation{Lawrence Livermore National Laboratory, Livermore, CA 94551, USA}
\author{M.~Guttormsen}
\affiliation{Department of Physics, University of Oslo, N-0316 Oslo, Norway}
\author{A.C.~Larsen}
\affiliation{Department of Physics, University of Oslo, N-0316 Oslo, Norway}
\author{G.E.~Mitchell} 
\affiliation{North Carolina State University, Raleigh, NC 27695, USA}
\affiliation{Triangle Universities Nuclear Laboratory, Durham, NC 27708, USA}
\author{J.~Rekstad}
\affiliation{Department of Physics, University of Oslo, N-0316 Oslo, Norway}
\author{A.~Schiller}
\affiliation{Department of Physics and Astronomy, Ohio University, Athens, OH 45701, USA}
\author{S.~Siem}
\affiliation{Department of Physics, University of Oslo, N-0316 Oslo, Norway}
\author{A.~Voinov}
\affiliation{Department of Physics and Astronomy, Ohio University, Athens, OH 45701, USA}

\begin{abstract} 
Nuclear level densities for $^{56,57}$Fe have been extracted from the
primary $\gamma$-ray spectra using ($^3$He,$^3$He$^{\prime}\gamma$)
and ($^3$He,$\alpha \gamma$) reactions.  Nuclear thermodynamic
properties for $^{56}$Fe and $^{57}$Fe  are investigated using
the experimental level densities.  These properties include entropy,
Helmholtz free energy, caloric curves, 
chemical potential, and heat capacity. In particular,
the breaking of Cooper pairs and single-quasiparticle entropy 
are discussed and shown to be important concepts for describing nuclear level density.  Microscopic model calculations are performed for level densities of $^{56,57}$Fe.  The experimental and calculated level densities are compared.  The average number of broken Cooper pairs and the parity distribution are extracted as a function of excitation energy for $^{56,57}$Fe from the model calculations.
\end{abstract}
\maketitle

\section{Introduction}
\label{sec:introduction}

Nuclear thermodynamics has attracted considerable attention in recent
years.  Several temperature-dependent nuclear properties such as 
nuclear shapes,  giant dipole resonance widths and their fluctuation
properties have been investigated in the literature \cite{kusnezov1998}.
In this context one of the most interesting topics is that of phase
transitions in atomic nuclei.  Different types of 
phase transitions have been 
discussed in nuclear physics.  One is the first order phase transition
in the multifragmentation of nuclei which occurs at high temperatures
\cite{dagostino1996}.  Negative heat capacities observed in
multifragmentation of nuclei have been obtained from energy
fluctuations and interpreted as an indicator of a first order phase
transition \cite{dagostino1996,chomaz2000}.

The second type of phase transition in atomic nuclei is the
transition from a phase with strong pairing correlations to a phase
with weak pairing correlations \cite{sano1963}.  The onset of a 
discontinuity in thermodynamic variables and the 
evolution of zeros of the
canonical \cite{sumaryada2007,schiller2002a} and grand-canonical
partition functions \cite{schiller2002a} in the complex plane have
been discussed  in terms of pairing transitions.

Recently \cite{kaneko2007}, thermal properties for
$^{56}$Fe have been calculated using both static-path approximation (SPA) and SPA 
plus random-phase approximation (RPA). These calculations show
 that the increase of the moment of inertia with
increasing temperature is correlated with the suppression of pairing
correlations.   

Furthermore, structures in the heat-capacity curve related to the
quenching of pairing correlations have been obtained within    
relativistic mean field theory \cite{agrawal2001}, the
finite-temperature Hartree-Fock-Bogoliubov theory \cite{egido2000},
and the 
shell-model Monte Carlo (SMMC) approach \cite{liu2001}.  An
$S$-shaped structure in the 
heat capacity curve derived from the level
densities of low-spin states has also been observed
experimentally \cite{schiller2001} and interpreted as a signature of a
pairing phase transition.

As a part of 
an ongoing effort, in the present work  we extract several
thermodynamic   properties of $^{56}$Fe and $^{57}$Fe isotopes
starting from nuclear level densities.  Details of the experiment, analysis tools, the Oslo method, and experimental level densities are presented in Sect. \ref{sec:experiment}.  
Thermodynamic concepts are discussed in
Sect. \ref{sec:thermo} and comparison with a combinatorial model is shown in Sect. \ref{sec:calc}.  Finally our results are summarized in Sect.
\ref{sec:conc}.

\section{Experimental details and data analysis}
\label{sec:experiment}
The self-supporting $^{57}$Fe target was bombarded by a 2-nA beam of
45-MeV $^3$He particles from the Oslo Cyclotron Laboratory at the
University of Oslo.  The target was isotopically enriched to 94.7\%,
and had a thickness of 3.4 mg/cm$^2$.  The outgoing charged particles were
recorded by eight Si $\triangle$E-E telescopes, which were collimated
and placed 5 cm away from the target at a ring located at 45$^{\circ}$ with
respect to the beam direction.  The particle telescopes covered 0.3\%
of the total solid angle.  The thicknesses of the front and the end
detectors were 140 and 3000 $\mu$m, respectively, and the particle
energy resolution was $\approx$ 0.3 MeV over the entire spectrum.  The
reaction $\gamma$ rays were detected by 28 collimated
5$^{\prime\prime}$ x 5$^{\prime\prime}$ NaI(Tl) detectors with a total
efficiency of $\approx$ 15\% of 4$\pi$, and with 6\% energy resolution
at 1.3 MeV.  In order to monitor the selectivity and populated spin
distribution of the reactions, one 60\% Ge(HP) detector was used.

The excitation energy of the 
final nucleus is determined from the known
$Q$ value and the reaction kinematics using     particle-$\gamma$
coincidence information.  A total $\gamma$-ray spectrum is obtained
for each excitation energy region.  These $\gamma$-ray spectra are
then unfolded using a Compton subtraction method
\cite{guttormsen1996}.  A two-dimensional primary $\gamma$-ray matrix
is extracted by applying a subtraction method to the unfolded
$\gamma$-ray spectra \cite{guttormsen1987}.  Basic assumptions and
details of the subtraction method are given in
Ref. \cite{guttormsen1987}.

The primary $\gamma$-ray matrix is factorized using the Brink-Axel
hypothesis \cite{brink,axel1962}, according to which the probability
of emitting a $\gamma$ ray from an excited state is proportional to
the $\gamma$-ray transmission coefficient $\mathcal T(E_{\gamma})$ and the
level density at the final energy $\rho (E-E_{\gamma})$.  This
factorization is determined by a least $\chi ^2$ method without
assuming any functional form for the level density and the
$\gamma$-ray transmission coefficient \cite{schiller2000}.
However, this method does not provide 
     a unique solution.  This factorization is                
invariant under the transformation \cite{schiller2000}

\begin{eqnarray}
{\tilde \rho}(E-E_{\gamma}) & = & A \exp(\alpha (E-E_{\gamma})) \rho(E-E_{\gamma})  \nonumber \\
{\tilde {\cal T}}(E_{\gamma}) & = & B \exp(\alpha E_{\gamma}) {\cal T}(E_{\gamma}),
\label{eq:trans}
\end{eqnarray}
where $A$, $B$, and $\alpha$ are the free parameters of the
transformation.  Therefore it is very important to determine
accurately the free parameters $A$, $B$, and $\alpha$ in order to find
the physical solution.  The parameters $A$ and $\alpha$ are determined
from the normalization of the level density to the discrete levels at
low excitation energies and to the density of the neutron resonances
at the neutron binding energy $B_n$.  The parameter $B$ is then
determined using the average total radiative width of neutron
resonances \cite{voinov2001}. 

Since there are no neutron resonance data for the $^{56}$Fe compound nucleus a different procedure for the
normalization of the level density is performed.  
We use $^{57}$Fe as a basis, since both average neutron resonance spacings $D$
and total radiative widths $\Gamma_{\gamma}$ of neutron resonances are well known \cite{mughabghab}. The
procedure for extracting the total level density $\rho(B_n)$ from the average resonance
spacing $D$ is described in Ref.~\cite{schiller2000}.
Then we apply the von Egidy and Bucurescu
parameterization of the back-shifted Fermi gas formula \cite{egidy2005}
\begin{equation}
\rho_{\mathrm BSFG}=\eta \frac{\exp(2\sqrt{aU})}{12\sqrt{2}a^{1/4}U^{5/4}\sigma},
\label{eq:bsfg}
\end{equation}
where $a$ is the level density parameter, and 
the intrisic excitation energy $U=E-E_1$ is determined by the backshift parameter $E_1$.
In this global description of the level density, shell corrections 
are taken into account in the estimation of the $a$ and $E_1$ parameters \cite{egidy2005}.
By fitting $\rho_{\mathrm BSFG}$ to $\rho(B_n)$ in $^{57}$Fe, 
we determine the normalization parameter $\eta = 0.852$ of Eq.~(\ref{eq:bsfg}). This value is then 
adopted for $^{56}$Fe together with the prescribed $a$ and $E_1$ parameters.
The parameters used are listed in Table \ref{ld_par}.

\begin{figure}[h!]
\begin{center}
\subfigure[]{
\includegraphics[scale=0.38]{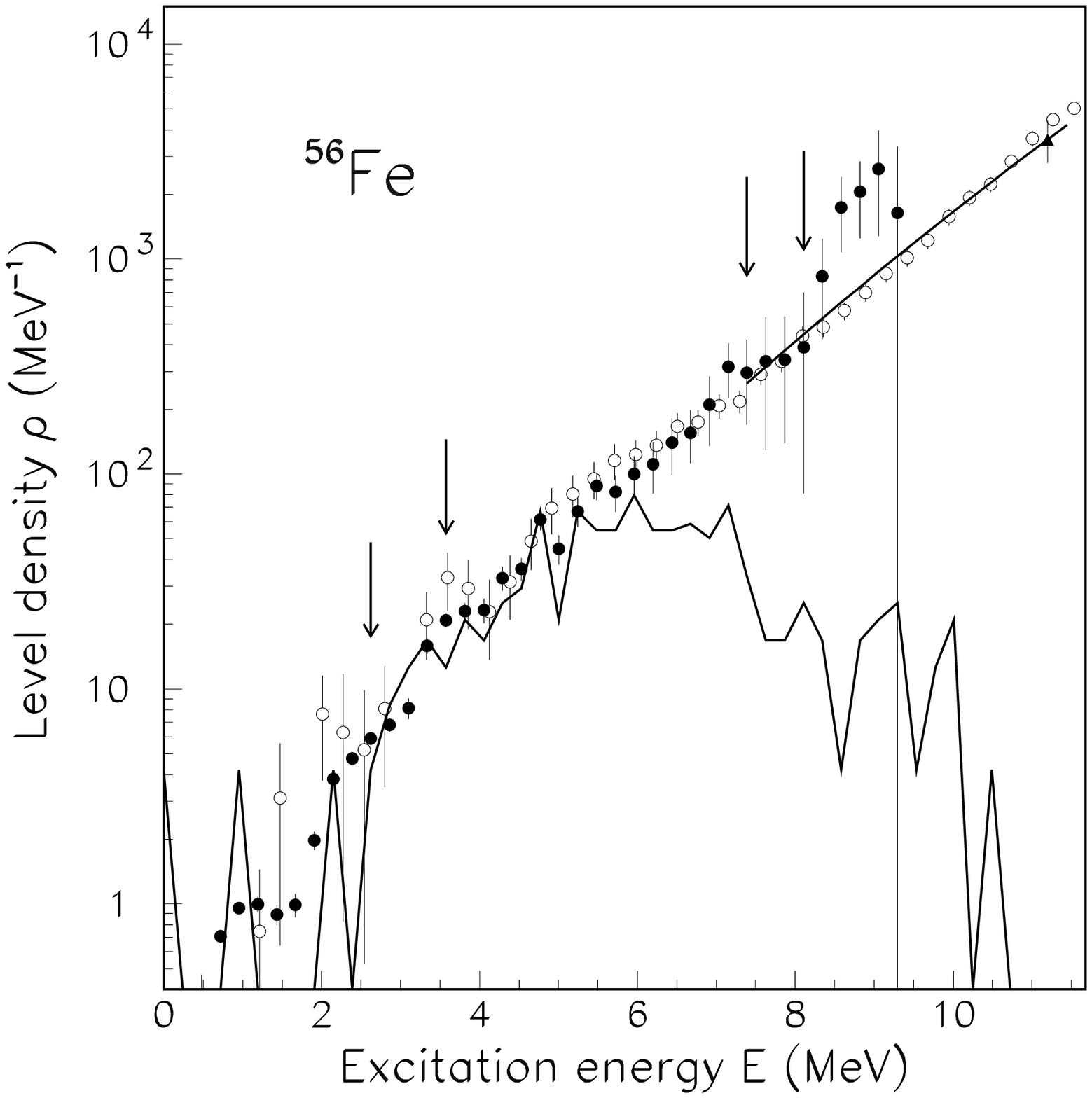}}
\subfigure[]{
\includegraphics[scale=0.38]{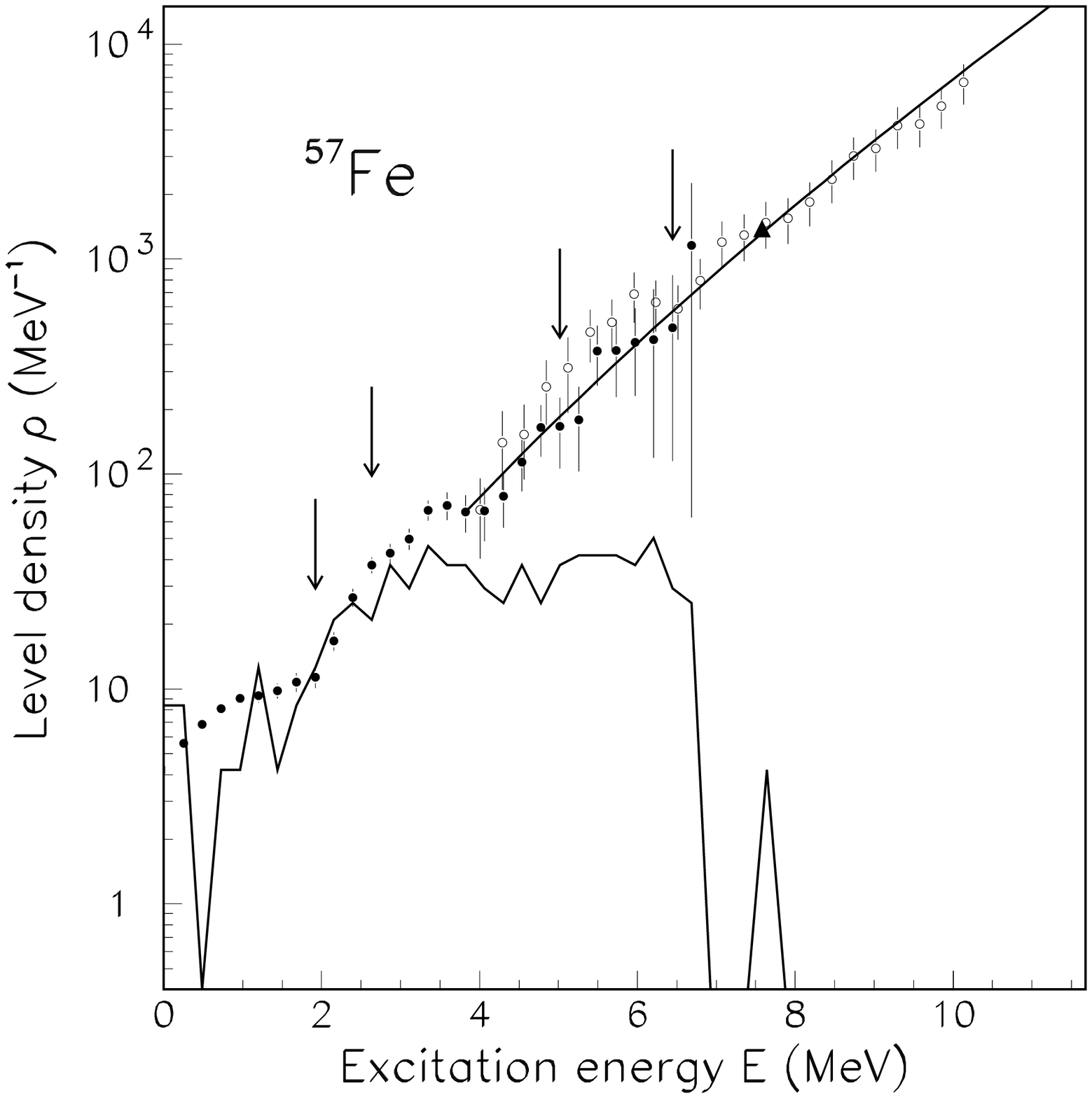}}
\caption{\label{fig:rho56_57} (a) Experimental level density 
of $^{56}$Fe (full circles).  
The jagged solid lines
represent the  level density obtained  from counting of discrete levels
\cite{firestone}.  The smooth solid curve is the renormalized level
density parametrization according to von Egidy and Bucurescu
\cite{egidy2005}.  The data points between the arrows at low and high
excitations are used for the normalization. The triangle at $B_n$ is determined from
$\rho_{\mathrm BSFG}$ with $\eta=0.852$.
The level density obtained from 
the $^{55}$Mn(d,n)$^{56}$Fe reaction
\cite{voinov2006} (open circles) is scaled with a factor of 1.3.  
 (b) Experimental level density of $^{57}$Fe (full circles). 
The triangle at $B_n$ is determined from neutron resonance spacings.
The level density obtained from the $^{58}$Fe($^{3}$He,$\alpha$)$^{57}$Fe reaction
\cite{voinov2007} (open circles) is scaled with a factor of 2.6.}
\end{center}
\end{figure}

Figure \ref{fig:rho56_57} shows the level
densities of $^{56,57}$Fe from the ground state up to $E$ $\sim$
$B_n-1$ MeV. They are normalized to discrete levels at
low excitation energies (jagged lines) and to the level densities at
the neutron binding energy $B_n$ (triangles). The arrows indicate the fitting regions used.
As described above, the triangles are determined from Eq.~(\ref{eq:bsfg}) with $\eta=0.852$
for $^{56}$Fe and from the observed $D$ value for $^{57}$Fe. 

The level density obtained with the Oslo method follows well that obtained by counting of discrete levels up to around 5 and 2.5 MeV
of excitation energies in $^{56,57}$Fe, respectively. Both nuclei
show an abrupt increase in level density at $E=2-3$ MeV, indicating the
first breaking of nucleon Cooper pairs.

Figure \ref{fig:rho56_57} also includes level densities from particle
evaporation data (open circles) measured with the $^{55}$Mn(d,n)$^{56}$Fe reaction \cite{voinov2006},
and with the $^{58}$Fe($^{3}$He,$\alpha$)$^{57}$Fe and
$^{59}$Co($d,\alpha$)$^{57}$Fe reactions \cite{voinov2007}.
The method determines the slope of the level densities, but
not the absolute normalization constant. Thus, these data 
have been scaled to match $\rho_{\mathrm BSFG}$ (solid smooth curves).
The slopes of the level densities from evaporation studies fit
very well with the employed $\rho_{\mathrm BSFG}$. This gives support to the 
level density parameters $a$ extracted as prescribed in Ref.~\cite{egidy2005}
and listed in Table~\ref{ld_par}.

\begin{figure*}[t!]
\includegraphics*[scale=0.6]{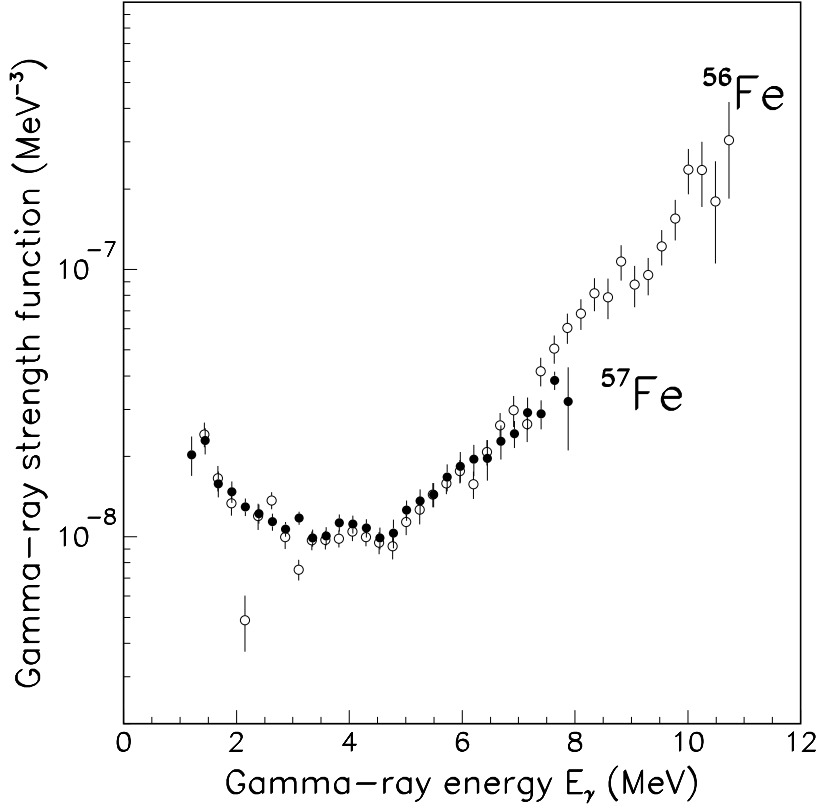}
\caption{Radiative strength functions for $^{56}$Fe and $^{57}$Fe.}
\label{fig:rsf56_57}
\end{figure*}

From Eq.~(\ref{eq:trans}) we see that the $\alpha$ parameter determines the slope of the transmission coefficient,
and thereby also the radiative strength function (RSF)  
$f(E_{\gamma}) = \mathcal T(E_{\gamma})/(2 \pi E_{\gamma}^3)$, assuming only dipole radiation. 
The RSFs have been published previously \cite{voinov2004,voinov2006}, but with
other normalization procedures. Thus, for completeness, we show in Fig.~\ref{fig:rsf56_57} the 
renormalized RSFs. In the $^{57}$Fe case, the $B$ parameter of Eq.~(\ref{eq:trans}) was determined from
the average radiative width $\langle \Gamma_{\gamma} \rangle $ of neutron resonances at $B_n$~\cite{mughabghab}.
For $^{56}$Fe, where no $\langle \Gamma_{\gamma} \rangle $ exists, we scale the total RSF to match
$^{57}$Fe. The parameters are listed in Table~\ref{ld_par}.

\begin{table}[hbtp]
\begin{center}
\parbox{14cm}{\caption{\label{ld_par}Back-shifted Fermi gas level density parameters and normalization constants.}}

\vspace{.1in}
\begin{tabular}{ccccccccc}\hline\hline
Isotope  &  $B_n$ &      a     &  $E_1$ & $D$        &$\sigma (B_n)$ & $\eta$   & $\rho(B_n)$  & $\langle \Gamma_{\gamma} \rangle $ \\ 
         &  (MeV) &(MeV$^{-1}$)&  (MeV) &(keV)       &               &          & (MeV$^{-1}$) & (meV)                              \\ \hline
$^{56}$Fe& 11.197 &  6.196     &  0.942 &2.6(6)$^a$  &4.049          & 0.852$^a$& 3600(800)$^a$& 2300(1200)$^a$                     \\
$^{57}$Fe&  7.646 &  6.581     & -0.523 & 22.0(17)   &3.834          & 0.852    & 1380(170)    & 900(470)                           \\ \hline\hline
\end{tabular}
\\$^a$ Based on systematics, see text.

\end{center}
\end{table}

\section{Thermodynamic quantities}
\label{sec:thermo}
Depending on the system under study, one can chose among different kinds of statistical ensembles in order to derive thermodynamic quantities. The thermodynamic quantities derived within different ensembles give the same results in the thermodynamic limit. On the other hand, the choice of a specific ensemble may change results significantly for small systems. For example, the caloric curves derived within the microcanonical and canonical ensembles coincide for large systems; but the two caloric curves depart from each other for small systems \cite{schiller2005,tavukcu}.

The microcanonical ensemble is commonly accepted as the appropriate
ensemble  to use in investigating atomic nuclei, as the nuclear force has a
short range and the nucleus does not share its excitation energy with
its surroundings.  However some thermodynamic quantities such as temperatures and heat capacities may have large fluctuations and negative values when derived within the microcanonical ensemble.  On the other hand, the canonical ensemble averages too much over structural changes of the system. Therefore, it is diffucult to chose an appropriate ensemble for a small system.  Here, we use both microcanonical and canonical ensembles to study the thermodynamic properties of the system.
 
\subsection{Microcanonical Ensemble}
The
microcanonical entropy is closely related to the
level density of the system at a given excitation energy.  Several
thermodynamic properties of the atomic nucleus can be derived from the
entropy.  The entropy is defined as the natural logarithm of the
multiplicity $\Omega$ of accessible states within the energy interval
$E$ and $E + \delta E$:
\begin{equation}
S(E)=k_B\ln \Omega(E),
\end{equation}
where $k_B$ is the Boltzmann's constant.  Here we set $k_B$ to unity so
that the entropy becomes dimensionless and thus the temperature $T$
has the unit of MeV.  One should note that the experimental level density is not the
true multiplicity of states; i.e., it does not include the $(2I+1)$
degeneracy of magnetic substates.  Thus, in order to obtain the state
density $\Omega$, one needs to know the spin distribution as a function of
excitation energy.  The spin distribution is usually assumed to be Gaussian with
a mean of $\langle2I+1\rangle=\sqrt{2\pi}\sigma$, where $\sigma$ is the spin
cut-off parameter which depends very weakly ($\sigma \propto E^{1/4}$)
on excitation energy.

However, in this work  we define a ``pseudo'' entropy based on the
experimental level density, i.e., $\Omega (E) = \rho (E)/\rho_0$.  The
normalization constant $\rho_0$ is introduced and adjusted such that
the third law of thermodynamics is fulfilled as $S(T\rightarrow 0)=0$.
The ground states of even-even nuclei represent a well-ordered system
with no thermal excitation and are characterized by zero entropy and
temperature.  Therefore the normalization constant is set to $\rho_0
=$ 1 MeV$^{-1}$ in order to obtain $S=\ln \Omega \sim 0$ in the ground state
band region of $^{56}$Fe.  The extracted $\rho_0$ is also used for the
$^{57}$Fe nucleus.

\begin{figure*}[t!]
\includegraphics*[scale=0.7]{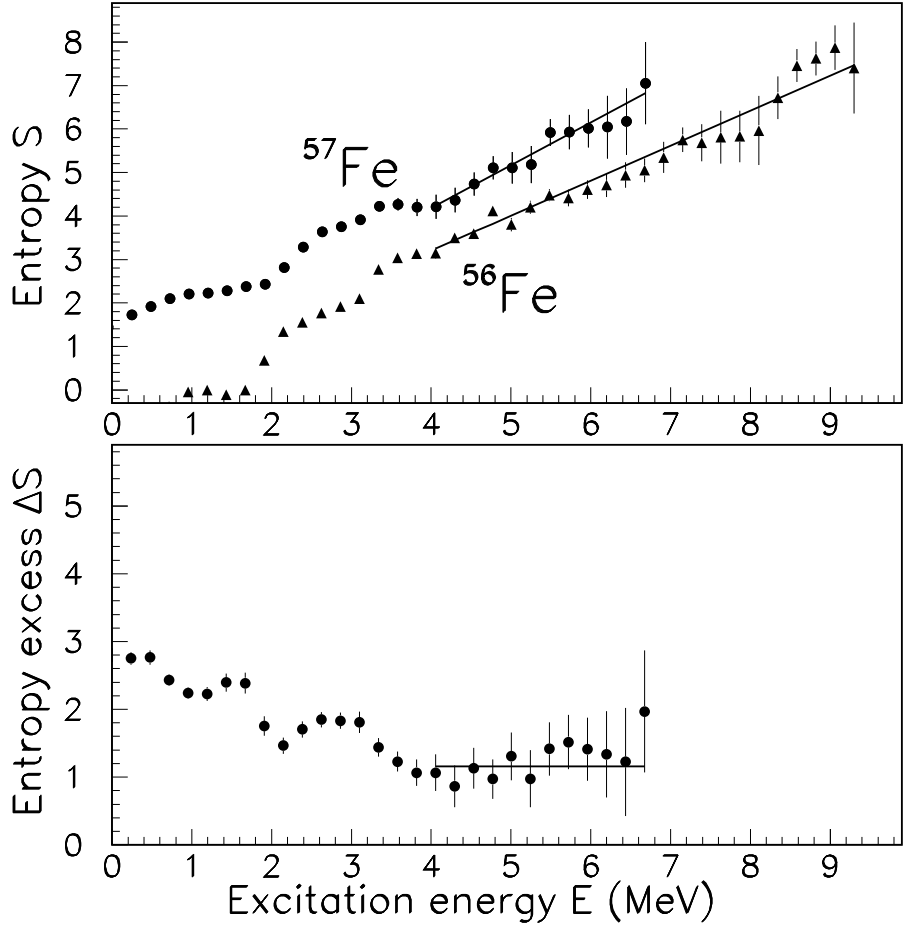}
\caption{Upper panel: Deduced entropies for $^{56,57}$Fe.  
The solid line is the constant temperature least square fit to the data.  
Lower panel: Deduced entropy excess for the single particle calculated 
using Eq.~(\ref{eq:entex}).  The entropy excess is $\Delta S=1.2(4)$.  }
\label{fig:feentr}
\end{figure*}

Figure \ref{fig:feentr} shows the entropies of $^{56,57}$Fe.  The
breaking of the first Cooper pair appears in the $E=2-3$ MeV excitation
energy region for both nuclei, indicating a slightly delayed
breaking for $^{57}$Fe due to the odd valence neutron.

The entropies of $^{56,57}$Fe appear fairly linear at high excitation
energies, and the slope of the entropy is related to the temperature of the system by 
\begin{equation}
T=(dS/dE)^{-1}_V.
\end{equation}
The entropies of $^{56,57}$Fe are fit with a constant temperature
model, shown with solid lines in Fig.~\ref{fig:feentr}.  From this
model, constant temperatures of $T=1.2(3)$ MeV and $T=1.0(3)$ MeV are
found for $^{56}$Fe and $^{57}$Fe, respectively.  These temperatures
are interpreted as the critical temperatures $T_c$ for the breaking of
nucleon pairs.

The entropy difference between the even-odd and even-even nuclei is
interpreted as the entropy of a single quasiparticle (particle or
hole), assuming that the entropy is an extensive (additive) quantity.
The entropy carried by the valence neutron particle in our case can be
estimated by

\begin{equation}
\Delta S=S(^{57}\text{Fe}) - S(^{56}\text{Fe}).
\label{eq:entex}
\end{equation}

\noindent
The 
lower panel of Fig. \ref{fig:feentr} shows the 
         single particle
entropy as a function of excitation energy.  The fluctuations at the
low excitation energies result from the lower pairing gap in the
odd-mass system.  At higher excitation energies above $\sim$ 4 MeV,
the entropy difference has a relatively constant value, revealing the 
statistical behavior of these isotopes, $\Delta S =1.2(4)$.  This
value is less than the value obtained for the rare-earth isotopes
which is $\Delta S =1.7$ \cite{guttormsen2001-2}.  This is
expected because $^{56}$Fe and $^{57}$Fe isotopes are in the vicinity
of closed shells, thus the entropy is less than that of the rare-earth
isotopes.

Assuming a constant $\Delta S$ and a constant
energy shift $\Delta$ between the two entropies, these two quantities
are connected to the critical temperature $T_c$
\cite{guttormsen2001-2} by
\begin{equation}
T_c = \frac{1}{\Delta S}\Delta.
\end{equation}
From this relation, we can calculate the entropy difference $\Delta S
=\Delta/T_c=1.5/1.1 \sim 1.4$, which is consistent with the
present observations.  Here, the gap parameter $\Delta$ is assumed to
include both the pairing gap and the contribution from the distance
between the Fermi surface and the closest neutron orbital.

\subsection{Canonical Ensemble}

In a canonical ensemble, a physical system is in thermal equilibrium with a heat reservoir at constant temperature and can exchange energy (but not particles) with the reservoir.  Analogous to the entropy in the microcanonical ensemble, the partition function is the starting point to obtain the thermodynamics of a system in the canonical ensemble.  The partition function for a given
temperature is determined by

\begin{equation}
Z(T)=\sum_{E=0}^{\infty} \rho(E)\delta E e^{-E/T},
\end{equation}

\noindent
where $\rho(E)$ is the measured level density, and $\delta E$ is the
energy bin used.  The summation in $Z(T)$ goes to infinity and our
level densities extend to $\sim B_n-1$ MeV.  Therefore we extrapolate
the experimental level densities using the BSFG model parameterized by
von Egidy and Bucurescu \cite{egidy2005}. The Helmholtz free energy
can be calculated from the partition function by

\begin{equation}
F(T)= -T \ln Z(T).
\end{equation}

\noindent
This equation gives the connection between statistical mechanics and
thermodynamics in the canonical ensemble, as does the entropy in the
microcanonical ensemble.  Then the entropy $S$, the average excitation
energy $\langle E \rangle$, the heat capacity $C_V$, and the
chemical potential $\mu$ at a given temperature can be derived from
$F(T)$ by

\begin{equation}
S(T)= -\left(\frac{\partial F}{\partial T}\right)_V,
\end{equation}

\begin{equation}
\langle E(T)\rangle =F+TS,
\end{equation}

\begin{equation}
C_V(T)= \left(\frac{\partial \langle E\rangle}{\partial T}\right)_V,
\end{equation}

\begin{equation}
\mu(T)= \frac{\partial F}{\partial n},
\end{equation}

\noindent
where $n$ is the number of thermal particles.  These thermal particles
outside a core of Cooper pairs are responsible for the thermal
properties of the nucleus at low excitations.  At higher temperatures,
pairing correlations are quenched, and the nucleus has a transition
from a paired to unpaired phase.

\begin{figure*}[t!]
\includegraphics*[scale=0.7]{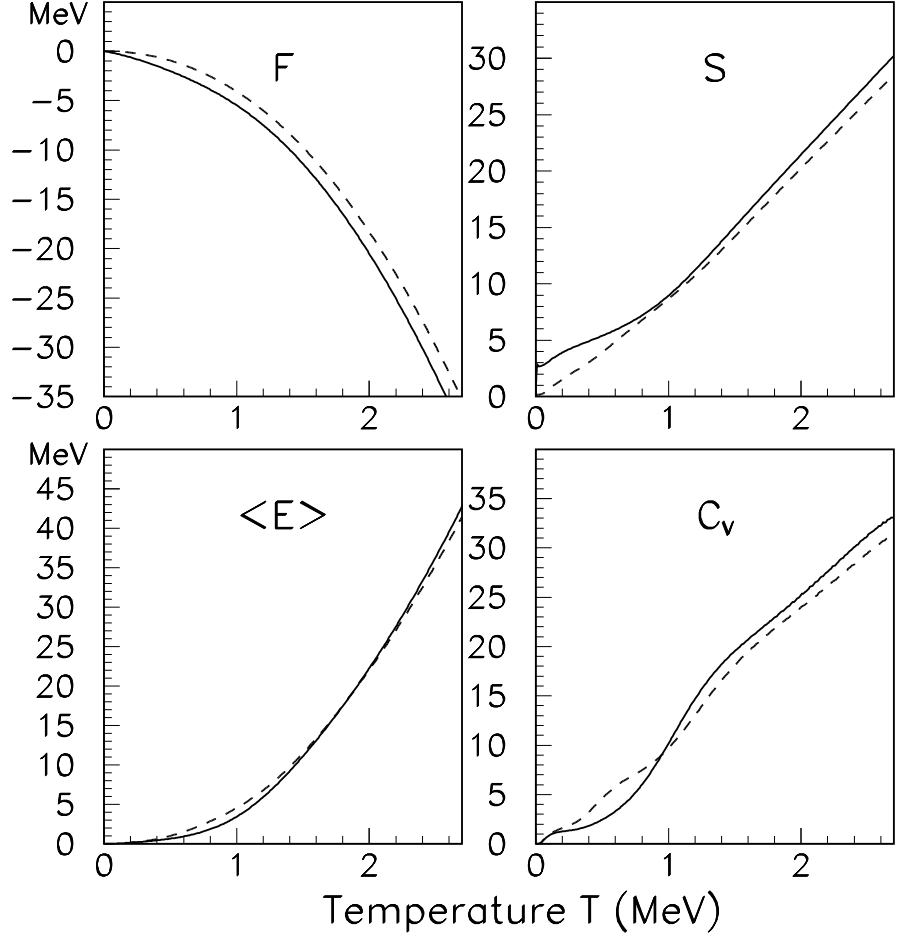}
\caption{The Helmholtz free energy $F$, average excitation energy $\langle E(T)\rangle$, 
entropy S, and heat capacity C$_V$ for $^{56}$Fe (dashed lines) and $^{57}$Fe (solid lines),
 deduced in the canonical ensemble.}
\label{fig:sfec}
\end{figure*}

\begin{figure*}[t!]
\includegraphics*[scale=0.7]{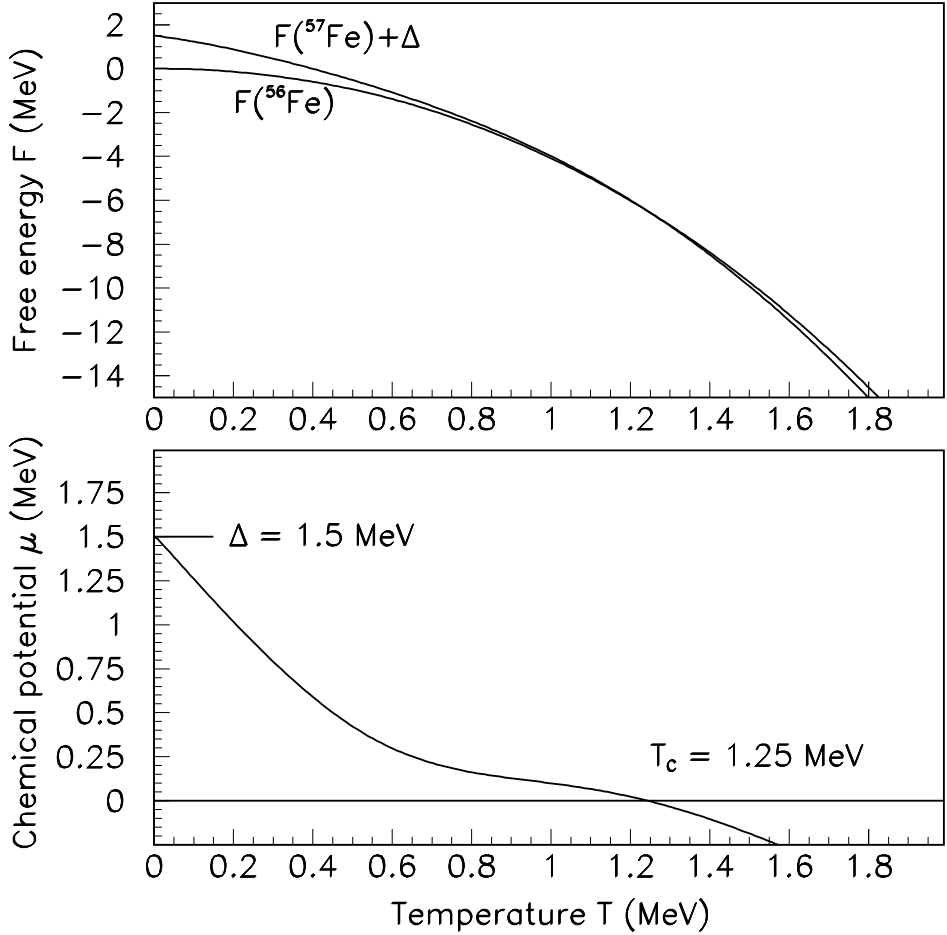}
\caption{The  experimental Helmholtz free energies deduced in the canonical ensemble for $^{56}$Fe and $^{57}$Fe.  
The critical temperature for the quenching of pairing correlations
 is indicated by $T_c$, where the chemical potential is $\mu \sim 0$.}
\label{fig:mu}
\end{figure*}
Figure \ref{fig:sfec} shows the Helmholtz free energy $F$, average
excitation energy $\langle E(T)\rangle$, entropy $S$, and heat capacity
C$_V$ for $^{56}$Fe (dashed lines) and $^{57}$Fe (solid lines).  The
Helmholtz free energy $F$ and the average excitation energy $\langle
E(T)\rangle$ behave smoothly as a function of temperature.  At around
$T \sim 1.3$ MeV, both isotopes are excited to energies comparable to
their respective neutron binding energies.  Average excitation
energies for $^{56}$Fe and $^{57}$Fe coincide only at one point $T\sim 2$ MeV.  
Above this temperature the odd isotope has larger values as the
temperature increases.  The entropy $S$ and heat capacity C$_V$ are the
first and second derivatives of $F$, respectively, and 
thus both reflect
thermal variations.  For temperatures below $T=1$ MeV, the entropy
difference for $^{56}$Fe and $^{57}$Fe reaches $\sim 2$, the entropy
of $^{57}$Fe being larger.  The entropies have similar values
for $T=1 - 1.5$ MeV.  Above $T=1.5$ MeV,  the 
two entropies start to
diverge.  Both heat capacities have an $S$ shape which is interpreted as
a fingerprint for pairing transitions in nuclei.  The structure in 
the heat capacity
C$_V$ for $^{57}$Fe is more pronounced than the C$_V$ for $^{56}$Fe, which
is the 
   opposite of what SMMC calculations predict.  The contribution to
the heat capacity from collective excitations is negligible, and has
no influence on the $S$ shape \cite{guttormsen2001}.

Figure \ref{fig:mu} shows the chemical potential ${\mu}$ 
which is defined as
the amount of energy required to excite a nucleon from the underlying
core of paired nucleons, while the entropy and volume held fixed.  The
typical energy cost for creating a quasiparticle is $- \Delta$ which
is also equal to the chemical potential.  
The chemical potential can be written as

\begin{equation}
\mu= \frac{\Delta F}{\Delta N}=\frac{F_{odd}-F_{even}}{1}=-\Delta .
\end{equation}

\noindent
thus giving $F_{even}=F_{odd}+\Delta$.  The energy curve of $^{57}$Fe
can be interpreted as the free energy for an even-even system with one
extra nucleon.

\subsection{Probability Density Functions}

An alternative way of studying the energy-temperature relation of a
system is the probability density function, which is the
probability of the system having energy $E$ for a given temperature,
and is given by

\begin{equation}
P(E)= \frac{\Omega(E) \exp(-E/T)}{Z(T)}.
\end{equation}

\begin{figure*}[t!]
\includegraphics*[scale=0.6]{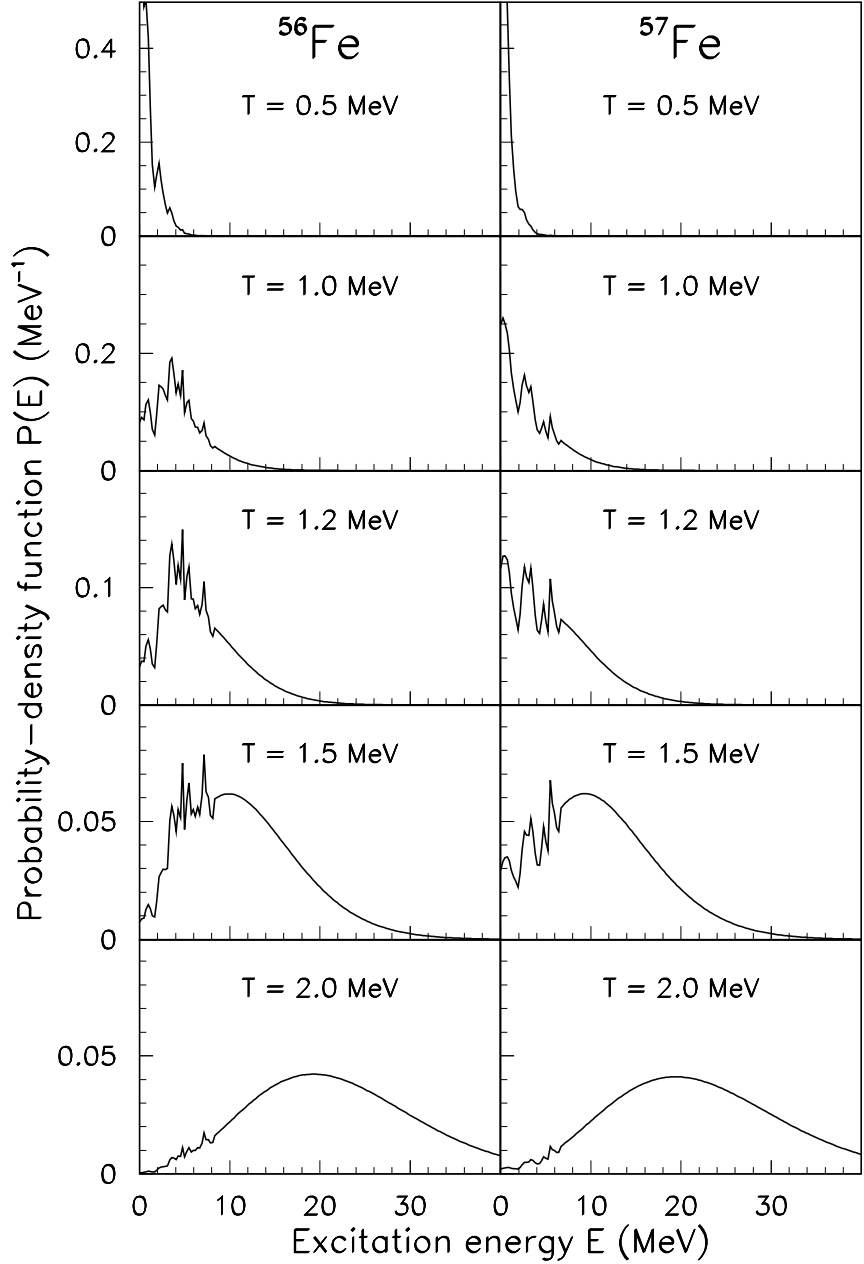}
\caption{Observed probability density functions for $^{56,57}$Fe for various temperatures.}
\label{fig:pet}
\end{figure*}

\noindent
In the thermodynamic limit $P(E)$ is very sharp - similar to a
$\delta$ function.  However, for small systems the multiplicity of
states is much smaller, which makes the probability distribution
broader.  Figure \ref{fig:pet} shows the probability density functions
for $^{56,57}$Fe nuclei for various temperatures.  Below $T \sim 1$
MeV, the distribution is mainly based on the experimental level
density.  For increasing temperatures, the energy of the nucleus
increases; therefore, the probability density function depends more
and more on the extrapolated level density.

Below $T=1.2$ MeV, the distribution is slightly broader for    
$^{56}$Fe due to     collective excitations below the pairing gap in
this nucleus.  The effect of critical temperatures obtained with the
constant temperature least-square fit can also be seen on the
probability density function.  At $T_c=1.2$ MeV and $T_c=1.0$ MeV for
$^{56,57}$Fe, respectively, the probability distribution becomes
broader where the breaking of nucleon pair process is strongest.

\section{Comparison of level densities with microscopic calculations}
\label{sec:calc}

In order to extract more information on the underlying nuclear structure resulting in the observed level density, we have performed microscopic calculations with the code \textsc{micro}~\cite{larsen_Sc}. The code is based on a model employing Bardeen-Cooper-Schrieffer (BCS) quasiparticles~\cite{BCS}, which are distributed on all possible proton and neutron configurations according to the given excitation energy $E$ of the nucleus. The advantage of this code is a fast algorithm that may include a large model space of single-particle states. Since level density is a gross property, the detailed knowledge of the many-particle matrix elements through large diagonalizing algorithms is not necessary.

The single-particle energies $e_{\rm sp}$ are calculated with the Nilsson Hamiltonian for an axially deformed core with a quadrupole deformation parameter $\epsilon_2$. The spin-orbit and centrifugal parameters $\kappa$ and $\mu$, together with the oscillator quantum energy $\hbar \omega_0 = 41 A^{-1/3}$ MeV between the harmonic oscillator shells, are also input to the code. 
Within the BCS model, the single-quasiparticle energies are defined by
\begin{equation}
e_{\rm qp}=\sqrt{(e_{\rm sp} - \lambda)^2 +\Delta^2 },
\end{equation}
where the Fermi level $\lambda$ is adjusted to reproduce the number of particles in the system and $\Delta$ is the pair-gap parameter.

\begin{figure*}[t!]
\centering
\includegraphics*[scale=0.6]{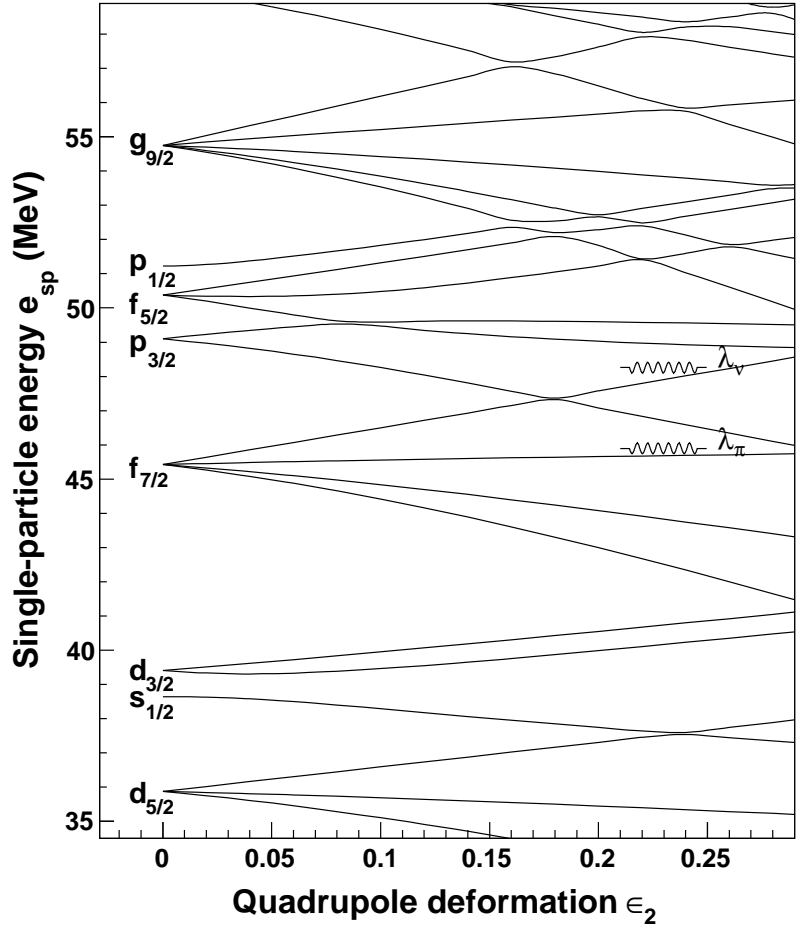}
\caption{Nilsson level scheme for $^{56}$Fe.}
\label{fig:nilsson}
\end{figure*}

In the calculations we have adopted the Nilsson parameters $\kappa=0.066$ and $\mu=0.32$ taken from Ref.~\cite{white}. The quadrupole deformation $\epsilon_2$ was set to 0.24 for $^{56}$Fe, and 0.25 for $^{57}$Fe~\cite{RIPL}. The Nilsson levels used in the calculations for $^{56}$Fe are shown in Fig.~\ref{fig:nilsson}, with the Fermi levels for the protons and neutrons. 

The rotational and vibrational terms, which are schematically added, contribute significantly to the total level density only in the lower excitation region. The rotational parameter $A_{\rm rot}$ was set to $0.12$ MeV in order to reproduce the ground-state rotational band of $^{56}$Fe~\cite{firestone}. The adopted pairing gap parameters $\Delta_{\pi}$ and $\Delta_{\nu}$ were evaluated from even-odd mass differences \cite{Wapstra} according to Ref.~\cite{BM} (see Table~\ref{tab:tab2} for a list of all parameters used).

\begin{table}\footnotesize
\caption{Input parameters for the model calculations.} 
\begin{tabular}{ccccccccc}
\hline
Isotope   &$\epsilon_2$&$\Delta_{\pi}$&$\Delta_{\nu}$& $A_{\rm rot}$ & $\hbar \omega _{0}$ & $\hbar \omega _{\rm vib}$ & $\lambda_{\pi}$ &  $\lambda_{\nu}$\\ 
          &            &      (MeV)   &     (MeV)    &      (MeV)    &  (MeV)  &       (MeV)  & (MeV) & (MeV)\\
\hline
$^{56}$Fe &     0.24   &  1.568       &  1.363       &    0.120      &   10.65 & 2.656       & 45.89 & 48.23  \\
$^{57}$Fe &     0.25   &  1.268       &  1.465$^a$       &    0.120      &   10.72 & 2.656       & 45.66 & 48.40    \\
\hline
\end{tabular}
\\
\label{tab:tab2}
$^a$ Taken from $^{58}$Fe. 
\end{table}

\begin{figure*}[t!]
\centering
\includegraphics*[scale=0.6]{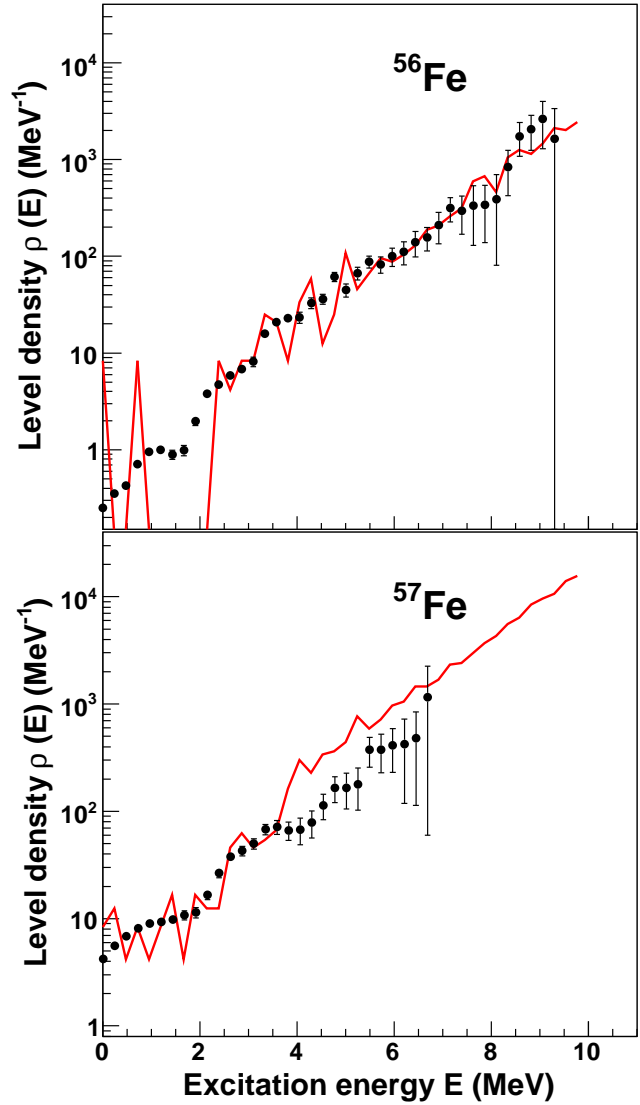}
\caption{Calculated level densities (solid line) compared with the experimental ones (data points with error bars) for $^{56,57}$Fe.}
\label{fig:micro}
\end{figure*}

The experimental and calculated level densities for $^{56,57}$Fe are shown in Fig.~\ref{fig:micro}. In general, there is a very good overall agreement between the measured level densities and the calculations, even without reducing the pairing gap. 
This is in contradiction with what is usually thought using the canonical ensemble. However, excitation energy around 8 MeV may be too low to see the overall quenching of the pairing correlations demonstrated in Figs.~(\ref{fig:sfec}) and (\ref{fig:mu}).  Microcanonical and canonical ensembles yield different results for small systems.  It is, therefore, no surprise that one measures different thermodynamic quantities within these ensembles.

In Fig.~\ref{fig:micro}, especially for $^{56}$Fe, it is gratifying how both the general functional form and the absolute magnitude of the level densities are reproduced. In the case of $^{57}$Fe, there is an overshoot in the calculated level density compared to the experimental data from about 3.5 MeV excitation energy. This excess could be due to the neutron $f_{5/2}$, $p_{1/2}$ and $g_{9/2}$ orbitals coming into play too soon, that is, they are too close to the neutron Fermi level. Another possibility is that the adopted (constant) deformation is too large at these excitation energies, where it may be that the nuclear structure favors a more spherical shape.

\begin{figure*}[t!]
\centering
\includegraphics*[scale=0.6]{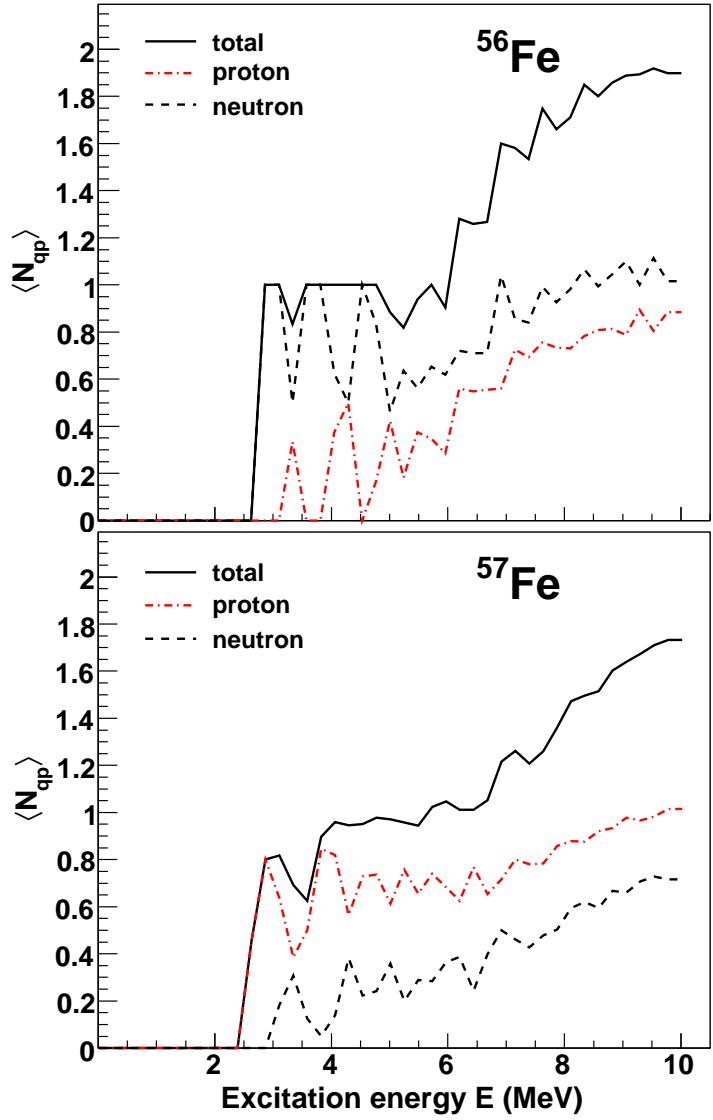}
\caption{The average number of broken Cooper pairs as a function of excitation energy for $^{56,57}$Fe.}
\label{fig:pairs}
\end{figure*}

Figure~\ref{fig:pairs} shows the average number of broken Cooper pairs $\langle N_{\mathrm{qp}}\rangle$ as a function of excitation energy. Both neutron and proton pairs are taken into account, adding up to the total number of broken pairs.  From Fig.~\ref{fig:pairs}, the pair-breaking process is seen to start at $E\approx 2.5$ MeV for both nuclei, in accordance with the values used for the pair-gap parameters (see Table~\ref{tab:tab2}). 

A prominent step structure at $\langle N_{\mathrm{qp}}\rangle = 1$ is seen clearly in $^{56}$Fe for excitation energies between $2.5 \leq E \leq 5.5$ MeV, and for $^{57}$Fe in the energy region $4.0 \leq E \leq 7.5$ MeV. This means that on average there is one Cooper pair broken, which could be either a proton or a neutron pair. For $^{56}$Fe, the total $\langle N_{\mathrm{qp}}\rangle$ exhibits an increase for $E > 5.5$ MeV up to about 9 MeV, where there seems to be a saturation for $\langle N_{\mathrm{qp}}\rangle \approx 2$. This is also reflected in the proton and neutron contributions to $\langle N_{\mathrm{qp}}\rangle$. A similar behavior can be seen in $^{57}$Fe for excitation energies higher than $\approx 8$ MeV. We observe that the onset of this increase is at higher excitation energies than for $^{56}$Fe. A possible explanation of this is that the $^{57}$Fe valence neutron blocks the quasi-particles created during the pair-breaking process, suppressing the average number of quasi-particles even at high excitation energies. This is supported by the fact that the neutron contribution to the total  $\langle N_{\mathrm{qp}}\rangle$ is lower or about the same as the proton contribution in $^{57}$Fe,  while in $^{56}$Fe, the neutron contribution is in general higher than the proton contribution.

The present model gives the opportunity of investigating the parity distribution as a function of excitation energy. For this purpose, we utilize the parity asymmetry defined in Ref.~\cite{gary} by
\begin{equation}
\alpha=\frac{\rho_+-\rho_-}{\rho_++\rho_-},
\end{equation}
which becomes $-1$ if only negative parity states are present, $+1$ if there are only positive parity states, and 0 when both parities are equally represented. In literature, one also find the expression $\rho_{-}/\rho_+$  which relates to $\alpha$ by

\begin{equation}
\frac{\rho_-}{\rho_+} = \frac{1-\alpha}{1+\alpha}.
\end{equation}

\begin{figure*}[t!]
\centering
\includegraphics*[scale=0.6]{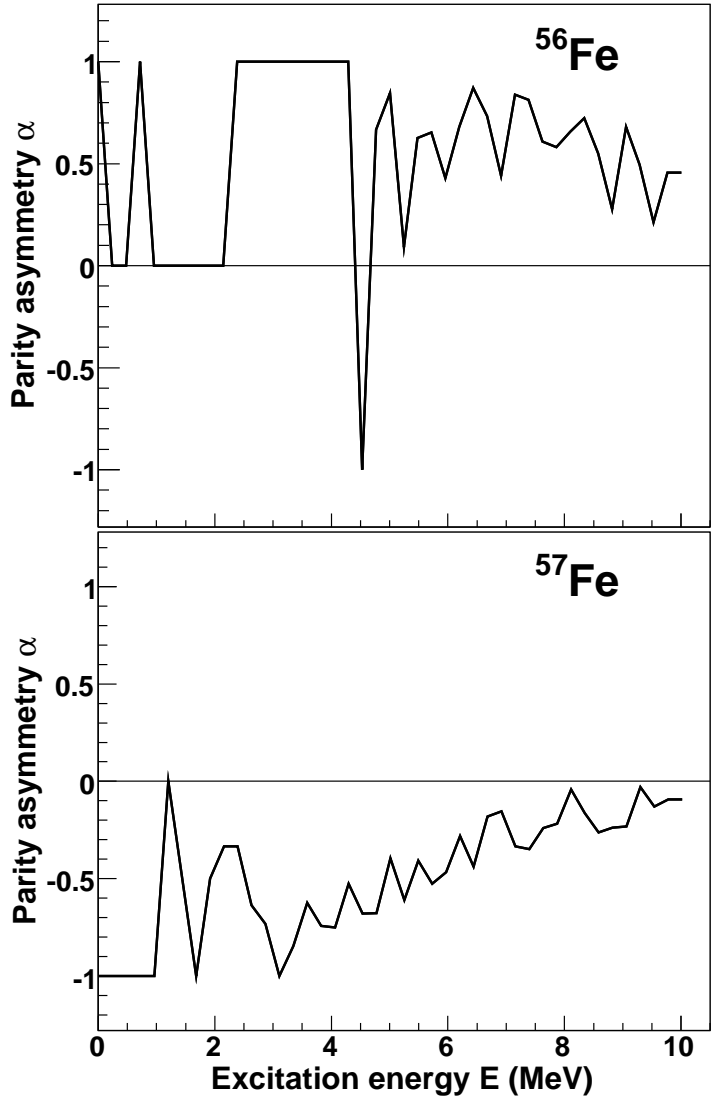}
\caption{The parity asymmetry as function of excitation energy for $^{56,57}$Fe.}
\label{fig:asymmetry}
\end{figure*}

\noindent
The parity asymmetry $\alpha$ for $^{56,57}$Fe is shown in Fig.~\ref{fig:asymmetry}. 


From Fig.~\ref{fig:asymmetry}, we see that there is an excess of positive parity states in $^{56}$Fe and negative parity states  in $^{57}$Fe. As the excitation energy increases, the parity asymmetry decreases for both nuclei, and shows an almost decay-like behavior, especially in the case of $^{57}$Fe. Here, the parity asymmetry is close to zero at $E = 9-10$ MeV. For $^{56}$Fe, there is still a significant overshoot of positive-parity states in this energy region, giving an average parity asymmetry of $\alpha \approx 0.35$ for $9.5 \simeq E \simeq 10$ MeV. 

We have compared our results with the shell-model Monte Carlo results of Y.~Alhassid \textit{et al.}~\cite{Alhassid}, and with recent macroscopic-microscopic calculations performed by D.~Mocelj \textit{et al.}~\cite{Rauscher}. In Fig.~2 of Ref.~\cite{Rauscher}, the ratio $\rho_{-}/\rho_+$ is shown for $^{56}$Fe, indicating a value of $\rho_{-}/\rho_+ \simeq 0.1$ at 10 MeV excitation energy. From Fig.~4 in Ref.~\cite{Alhassid}, the ratio $\rho_{-}/\rho_+ \simeq 0.2$ for $E=10$ MeV. 
Using an average parity asymmetry of $\alpha \approx 0.35$ as determined in the previous section, we obtain $\rho_{-}/\rho_+ \approx 0.5$, which implies a considerable number of additional positive-parity states than predicted by the macroscopic-microscopic model and the shell-model Monte Carlo approach. For $^{57}$Fe with $\alpha \approx 0.1$ for $9.5 \simeq E \simeq 10$ MeV, we find $\rho_{-}/\rho_+ \approx 0.8$. However, one should note that in our calculations, the parity asymmetry strongly depends on the position of the neutron $g_{9/2}$ orbital relative to the Fermi level (see Fig.~\ref{fig:nilsson}).

\section{Summary}
\label{sec:conc}
Nuclear level densities for $^{56,57}$Fe are renormalized using the
new level density parameterization suggested by von Egidy and
Bucurescu.  The level densities obtained with the Oslo method agree
well with those obtained from other experiments.  The experimental
level densities are used to extract thermodynamic quantities.  The
entropies for $^{56,57}$Fe obtained in the 
   microcanonical ensemble reveal
step structures indicating the breaking of     nucleon Cooper pairs.
The entropy carried by the single neutron at higher excitation energies (4~MeV $< E <$ 7~MeV) is estimated to be $\Delta S
= 1.2(4) k_B$ which is smaller than that of the rare-earth isotopes.

Assuming a constant $\Delta S$, a critical temperature for the
depairing process was determined.
      In the       canonical ensemble, several
thermodynamic properties were 
    investigated.  Probability density
functions for $^{56,57}$Fe were also extracted which reveal the
difference between small systems such as 
     the atomic nucleus and a
system in the thermodynamic   limit.

Microscopic model calculations based on BCS quasiparticles were performed.  The overall agreement between the experimental and calculated level densities are good.  Step structures observed in the experimental level densities are also observed in the plot of the average number of broken pairs as a function of excitation energy.  The parity distributions obtained from model calculations for $^{56,57}$Fe indicate a decrease of parity asymmetry for both isotopes as the excitation energy increases. The microscopic model, which is expressed within the microcanonical ensemble (fixed $E$ with no heat bath), describes the observed level densities without the need of attenuation of the pairing-gap energies for excitation energies below 8 MeV.

\begin{acknowledgments}

This work was supported in part by the U.S. Department of Energy under
grants number DE-FG02-97-ER41042 and DE-FG52-06NA26194.  In addition, this work was performed under the auspices of
the U.S. Department of Energy by the University of California,
Lawrence Livermore National Laboratory under contract
No. W-7405-ENG-48.  Financial support from the Norwegian Research
Council (NFR) is gratefully acknowledged.

\end{acknowledgments}


\end{document}